\documentclass{PoS}

\usepackage[utf8x]{inputenc}

\usepackage{amssymb}
\usepackage{amsmath}

\newcommand{\rt}{{\mathbf{r}}}
\newcommand{\xt}{{\mathbf{x}}}
\newcommand{\yt}{{\mathbf{y}}}
\newcommand{\bt}{{\mathbf{b}}}
\newcommand{\kt}{{\mathbf{k}}}
\newcommand{\qt}{{\mathbf{q}}}
\newcommand{\lt}{\mathbf{l}}
\newcommand{\ktt}{k_\perp} % scalar
\newcommand{\qtt}{q_\perp} % scalar
\newcommand{\ltt}{l_\perp} % scalar

\newcommand{\ud}{\mathrm{d}}
\newcommand{\tr}{\, \mathrm{Tr} \, }
\newcommand{\nc}{{N_\mathrm{c}}}
\newcommand{\cf}{C_\mathrm{F}}

\newcommand{\qso}{Q_\mathrm{s,0}}
\newcommand{\lqcd}{\Lambda_{\mathrm{QCD}}}
\newcommand{\as}{\alpha_{\mathrm{s}}}

\newcommand{\ical}{\mathcal{I}}
\newcommand{\jcal}{\mathcal{J}}
\newcommand{\scal}{\mathcal{S}}
\newcommand{\kcal}{\mathcal{K}}

\def\figscale{1.31}
\def\figspace{1mm}

\title{Implementing consistent NLO factorization in single inclusive forward hadron production}

\ShortTitle{Implementing consistent NLO factorization in single inclusive forward hadron production}

\author{\speaker{B. Duclou\'e}\\
        Department of Physics, P.O. Box 35, 40014 University of Jyväskylä, Finland\\
        and \\
        Helsinki Institute of Physics, P.O. Box 64, 00014 University of Helsinki,
        Finland \\
        E-mail: \email{bertrand.b.ducloue@jyu.fi}}
    
\author{T. Lappi\\
        Department of Physics, P.O. Box 35, 40014 University of Jyväskylä, Finland\\
        and \\
        Helsinki Institute of Physics, P.O. Box 64, 00014 University of Helsinki,
        Finland \\
    	E-mail: \email{tuomas.v.v.lappi@jyu.fi}}
    
\author{Y. Zhu\\
        Department of Physics, P.O. Box 35, 40014 University of Jyväskylä, Finland\\
        and \\
        Helsinki Institute of Physics, P.O. Box 64, 00014 University of Helsinki,
        Finland \\
    	E-mail: \email{yan.zhu@jyu.fi}}

\abstract{Single inclusive forward hadron production in high-energy hadron collisions can provide an important test of the Color Glass Condensate picture at small $x$. Recent studies of this process at next-to-leading order have led to problematic results, with cross sections becoming negative at large transverse momenta. We study a new formulation of this quantity proposed recently by Iancu et al. We show that it leads to physical results up to large transverse momenta at fixed coupling. Taking into account running coupling effects in a way that is consistent with existing DIS calculations still poses a challenge.}

\FullConference{XXV International Workshop on Deep-Inelastic Scattering and Related Subjects\\
		3-7 April 2017\\
		University of Birmingham, UK}
		
\allowdisplaybreaks

\begin{document}

\section{Introduction}

Forward particle production in high-energy hadron collisions can provide valuable information on the small-$x$ behavior of parton densities in the target. In this regime, the gluon density is expected to saturate due to nonlinear recombination effects and to evolve according to the Balitsky-Kovchegov (BK) equation~\cite{Balitsky:1995ub,Kovchegov:1999yj}. Many studies have been performed in this formalism at leading order (LO) accuracy. Recently progress has been made to extend this formalism to next-to-leading order (NLO) accuracy. In particular, the cross section for single inclusive hadron production has been calculated at NLO in Refs.~\cite{Chirilli:2011km,Chirilli:2012jd}. Unfortunately, the first numerical evaluation of these expressions showed that at large transverse momentum the NLO corrections are large and negative, making the total NLO cross section negative~\cite{Stasto:2013cha}. Many works have been devoted to understanding this issue~\cite{Kang:2014lha,Altinoluk:2014eka,Watanabe:2015tja,Ducloue:2016shw}.
Recently a new formulation of the NLO cross section, leading to explicitly positive cross sections, was proposed~\cite{Iancu:2016vyg}. The goal of the present work is to present a practical numerical study of this formulation and compare it to the previously used ``CXY'' formulation.

\section{Expressions for the NLO cross section}

In this work we focus on the $q \to q$ channel, which exhibits the same general features as the total cross section~\cite{Ducloue:2016shw}. In addition we leave out the fragmentation functions which do not affect our discussion but would make the numerical implementation more cumbersome. Following the notations used in \cite{Ducloue:2016shw}, we write the unsubtracted CXY quark multiplicity as~\cite{Chirilli:2011km,Chirilli:2012jd}
\begin{align}\label{eq:nlosigma}
 \frac{\ud N^{pA \to qX}}{\ud^2\kt \ud y}
= & x_p q(x_p) \frac{\scal_0(\ktt) }{(2\pi)^2} \nonumber \\
& + \frac{\as}{2\pi^2}
\int_{x_p}^{\xi_\text{max}} \ud \xi \frac{1+\xi^2}{1-\xi}
\frac{x_p}{\xi} q\left(\frac{x_p}{\xi}\right) \left\{\cf \ical(\ktt,\xi,X(\xi)) + \frac{\nc}{2}\jcal(\ktt,\xi,X(\xi)) \right\} \nonumber \\
& - \frac{\as}{2\pi^2}
\int_{0}^{\xi_\text{max}} \ud \xi \frac{1+\xi^2}{1-\xi}
x_p q\left(x_p \right) \left\{\cf \ical_v(\ktt,\xi,X(\xi)) + \frac{\nc}{2}\jcal_v(\ktt,\xi,X(\xi)) \right\} ,
\end{align}
with
\begin{eqnarray}
\ical(\ktt,\xi,X(\xi)) \!\! &=& \!\!
\int \frac{\ud^2 \qt}{(2\pi)^2} 
\left[\frac{\kt-\qt}{(\kt-\qt)^2} - \frac{\kt-\xi \qt}{(\kt-\xi \qt)^2} \right]^2
\scal(\qtt,X(\xi)) \, ,
\\ 
% ******************
\label{eq:defJ}
\jcal(\ktt,\xi,X(\xi)) \!\! &=& \!\!
\int \frac{\ud^2 \qt}{(2\pi)^2} 
\frac{2(\kt-\xi\qt)\cdot(\kt-\qt)}{(\kt-\xi\qt)^2(\kt-\qt)^2}
\scal(\qtt,X(\xi)) \nonumber \\
&& -\int \frac{\ud^2 \qt}{(2\pi)^2} \frac{ \ud^2\lt}{(2\pi)^2}
\frac{2(\kt-\xi\qt)\cdot(\kt-\lt)}{(\kt-\xi\qt)^2(\kt-\lt)^2}
\scal(\qtt,X(\xi))\scal(\ltt,X(\xi)) \, ,
\\
% ******************
\ical_v(\ktt,\xi,X(\xi)) \!\! &=& \!\!
\int \frac{ \ud^2 \qt }{(2\pi)^2}
\left[\frac{\kt-\qt}{(\kt-\qt)^2} - \frac{\xi\kt-\qt}{(\xi \kt-\qt)^2} \right]^2
\scal(\ktt,X(\xi)) \, ,
\\
% ******************
\label{eq:defJv}
\jcal_v(\ktt,\xi,X(\xi)) \!\! &=& \!\!
\left[
\int \frac{\ud^2 \qt}{(2\pi)^2} 
\frac{2(\xi\kt-\qt)\cdot(\kt-\qt)}{(\xi\kt-\qt)^2(\kt-\qt)^2} \right. \nonumber \\
&& \left. - \int \frac{\ud^2 \qt}{(2\pi)^2} \frac{  \ud^2\lt}{(2\pi)^2}
\frac{2(\xi\kt-\qt)\cdot(\lt-\qt)}{(\xi\kt-\qt)^2(\lt-\qt)^2}
\scal(\ltt,X(\xi))
\right]
\scal(\ktt,X(\xi))
.
\end{eqnarray}
These expressions involve $\scal$, the Fourier-transform of the dipole correlator
\begin{equation}
\scal(\ktt)=\scal(\ktt,\bt)=\int \ud^2\rt e^{-i\kt\cdot\rt} S(\rt) \, , \quad
S(\rt=\xt-\yt)=\left< \frac{1}{\nc}\tr V(\xt)V^\dag(\yt) \right>,
\end{equation}
where $V(\xt)$ is a fundamental representation Wilson line in the target's color field.

The kinematical variables appearing in Eqs.~(\ref{eq:nlosigma})-(\ref{eq:defJv}) are $x_p=\ktt e^{y}/\sqrt{s}$, $x_g=\ktt e^{-y}/\sqrt{s}$ and $\ktt=|\kt|$. The variable $\xi$ is the longitudinal momentum fraction of the incoming quark taken by the fragmenting one, i.e. the incoming quark carries a fraction $x_p/\xi$ of the projectile proton's momentum. $x_p$ and $x_g$ correspond to the longitudinal momentum fractions probed at leading order in the projectile and the target, respectively.

So far we have not specified the rapidity dependence $X(\xi)$ of the dipole correlators appearing in these expressions. In Ref.~\cite{Stasto:2013cha} this dependence is taken as $X(\xi)=x_g$, which is the longitudinal momentum fraction probed in the target at leading order. On the contrary, the discussion in Ref.~\cite{Iancu:2016vyg} leads to a dependence of $X$ on the kinematics of the radiated gluon at NLO, according to $X(\xi)\approx x_g/(1-\xi)$ in the usual ``Regge'' kinematics where it is assumed that all transverse momenta are of the same order. As will be shown in the following, the difference between the two choices $X(\xi)=x_g$ and $X(\xi)=x_g/(1-\xi)$ becomes important at large transverse momentum.

\subsection{$\nc$-terms}

We first consider the NLO corrections proportional to the $\nc$ color factor in Eq.~(\ref{eq:nlosigma}). It was shown in Ref.~\cite{Ducloue:2016shw} that these corrections are the ones leading to negative cross sections at large transverse momentum in Ref.~\cite{Stasto:2013cha}. One can write the sum of the leading order contribution and the $\nc$ corrections as
\begin{equation}\label{eq:nc_init}
\frac{\ud N^{\text{LO}+\nc}}{\ud^2\kt \ud y}
= x_p q(x_p) \frac{\scal_0(\ktt)}{(2\pi)^2}
+ \as\int_0^{1-x_g/x_0} \frac{\ud \xi}{1-\xi} \kcal(\ktt,\xi,X(\xi)) ,
\end{equation}
where we introduced the function $\kcal$, defined as
\begin{equation}
\kcal\!(\ktt,\xi,X(\xi))\!=\!\frac{\nc}{(2\pi)^2}(1\!+\xi^2) \!\Big[\!\theta(\xi\!-\!x_p)\frac{x_p}{\xi} q\!\left(\!\!\frac{x_p}{\xi}\!\!\right) \!\!\jcal(\ktt,\xi,X(\xi)) -x_p q\left(x_p \right)
\!\!\jcal_v(\ktt,\xi,X(\xi)) \!\Big].
\end{equation}
In Eq.~(\ref{eq:nc_init}), the correlator $\scal_0$ in the lowest order contribution corresponds to an unevolved target and the upper limit on the $\xi$ integral ensures that the target is not probed at values of $X=x_g/(1-\xi)$ larger than the initial condition $x_0$. Since $\kcal(\ktt,\xi,X)$ does not vanish when $\xi \to 1$ at fixed $X$, the integral over $\xi$ in the cross section develops a large logarithm at small $x_g$, which should be resummed in the Balitsky-Kovchegov evolution of the target. One can then identify $\scal_0$ with the initial condition for this evolution, at the initial rapidity $\ln(1/x_0)$. This allows to rewrite Eq.~(\ref{eq:nc_init}) as
\begin{equation}\label{eq:nc_unsub}
\frac{\ud N^{\text{LO}+\nc}}{\ud^2\kt \ud y}
= x_p q(x_p) \frac{\scal(\ktt,x_0)}{(2\pi)^2} + \as
\int_0^{1-x_g/x_0} \frac{\ud \xi}{1-\xi} \kcal(\ktt,\xi,X(\xi)) \, ,
\end{equation}
which is explicitly positive at all transverse momenta as long as the initial condition is. Noting that using the expressions of $\jcal$ and $\jcal_v$ one can write the BK equation in its integral form,
\begin{equation}\label{eq:integralBK}
\scal(\ktt,x_g)=\scal(\ktt,x_0)+2 \as \nc \int_{0}^{1-x_g/x_0} \frac{\ud \xi}{1-\xi} 
\left[\jcal(\ktt,1,X(\xi)) - \jcal_v(\ktt,1,X(\xi))\right],
\end{equation}
we can also rewrite Eq.~(\ref{eq:nc_unsub}) as
\begin{equation}\label{eq:nc_sub}
\frac{\ud N^{\text{LO}+\nc}}{\ud^2\kt \ud y}
 = x_p q(x_p) \frac{\scal(\ktt,x_g)}{(2\pi)^2}+\as
\int_0^{1-x_g/x_0} \frac{\ud \xi}{1-\xi} \left[\kcal(\ktt,\xi,X(\xi)) - \kcal(\ktt,1,X(\xi))\right] \, ,
\end{equation}
which is strictly equivalent and thus also positive. In the following we will refer to the formulation~(\ref{eq:nc_unsub}) as the ``unsubtracted'' version of the cross section and~(\ref{eq:nc_sub}) as the ``subtracted'' version. 

Let us now discuss the relation of Eqs.~(\ref{eq:nc_unsub}) and~(\ref{eq:nc_sub}) to the CXY formulation studied in Ref.~\cite{Stasto:2013cha}. This relation is most easily obtained from the ``subtracted'' formulation~(\ref{eq:nc_sub}), by replacing the rapidity of the dipole correlators by $X(\xi)=x_g$ and the upper limit of the $\xi$ integration by 1. This approximation is justified by considering that, because of the subtraction of $\kcal(\ktt,1,X(\xi))$, the integral over $\xi$ in Eq.~(\ref{eq:nc_sub}) should be dominated by the region where $\xi$ is close to 0.

\subsection{$\cf$-terms}

We now consider the NLO corrections proportional to $\cf$ in Eq.~(\ref{eq:nlosigma}). In Ref.~\cite{Ducloue:2016shw} it was shown that these corrections are positive and thus do not contribute to the negativity problem at large transverse momentum. However these terms present collinear divergences that have to be absorbed into the DGLAP evolution of the projectile's quark distributions and in the fragmentation functions. After subtracting the corresponding $1/\varepsilon$ poles, we can write the $\cf$-terms as
\begin{align}\label{eq:cf}
\frac{\ud N^{\cf}}{\ud^2\kt \ud y} \equiv \frac{\as}{2\pi^2} \cf
\bigg[ & \int_{x_p}^{1-x_g/x_0} \ud \xi \frac{1+\xi^2}{1-\xi} 
\frac{x_p}{\xi} q\left(\frac{x_p}{\xi}\right) \ical^\text{finite}(\ktt,\xi,X(\xi)) \nonumber \\
& - \int_{0}^{1-x_g/x_0} \ud \xi \frac{1+\xi^2}{1-\xi}
x_p q\left(x_p \right) \ical_v^\text{finite}(\ktt,\xi,X(\xi)) \bigg] ,
\end{align}
with
\begin{align}
\label{eq:Ifinite}
\ical^\text{finite}(\ktt,\xi,X(\xi)) =& \int \frac{\ud^2\rt}{4\pi} S(\rt,X(\xi)) \ln \frac{c_{0}^{2}}{\rt^{2}\mu ^{2}}\left( e^{-i\kt\cdot \rt}+\frac{1}{\xi^2}e^{-i\frac{\kt}{\xi }\cdot \rt}\right) \nonumber \\
& - 2 \int \frac{\ud^2 \qt}{(2\pi)^2} \frac{(\kt-\xi\qt)\cdot(\kt-\qt)}{(\kt-\xi\qt)^2(\kt-\qt)^2} \scal(\qtt,X(\xi)) \, ,\\
\label{eq:Ivfinite}
\ical_v^\text{finite}(\ktt,\xi,X(\xi)) =& \frac{\scal(\ktt,X(\xi))}{2\pi} \left(\ln{\frac{\ktt^2}{\mu^2}}+ \ln(1-\xi)^2\right).
\end{align}
In the case of the $\nc$-terms, the choice $X=x_g/(1-\xi)$ was motivated by the relation to the integral BK equation. Since the $\cf$-terms are not related to the rapidity evolution of the target, one could in principle choose another scale here. However, here we choose to evaluate the dipole correlators at the same rapidity in the $\cf$ and $\nc$ terms, since doing otherwise would be quite unnatural. This leads to the same limit $\xi<1-x_g/x_0$ on the $\xi$ integral.

\section{Results}

\begin{figure*}
	\includegraphics[scale=\figscale]{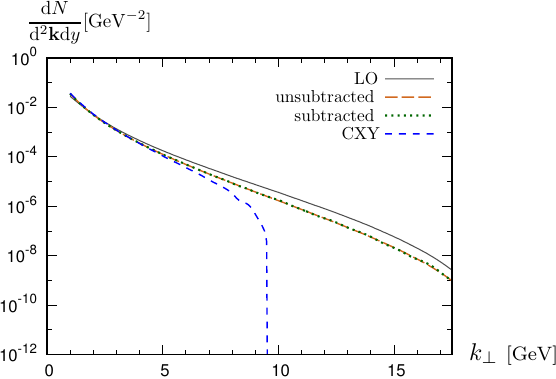}
	\hspace{\figspace}
	\includegraphics[scale=\figscale]{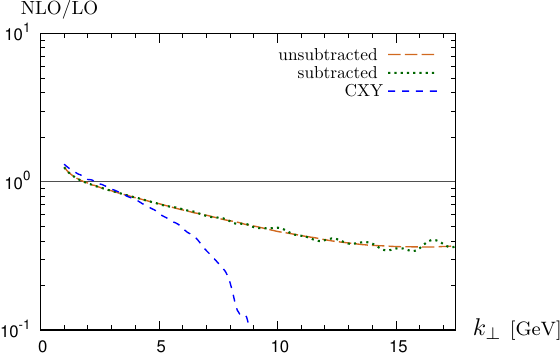}
	\caption{Results at fixed coupling for $\sqrt{s}=500$ GeV and $y=3.2$, using the unsubtracted and subtracted expressions or the CXY approximation. Left: multiplicity. Right: NLO/LO ratio.}
	\label{fig:fcBK_NLO}
\end{figure*}

Let us now turn to our numerical results. For simplicity we consider only contributions from up quarks in the projectile, and their distribution is obtained from the MSTW2008 NLO parametrization~\cite{Martin:2009iq}. We choose the factorization scale $Q=\ktt$, the center of mass energy $\sqrt{s}=500$ GeV and the rapidity of the produced quark $y=3.2$. We use a fixed value for the strong coupling, $\as=0.2$, both in the expression of the cross section and when solving the LO BK equation which provides the rapidity evolution of the dipole correlators. The initial condition for this evolution is taken according to the MV parametrization~\cite{McLerran:1993ni},
\begin{equation}
S(\rt,x_0)=\exp\left[-\frac{\rt^2 \qso^2}{4}\ln{\left(\frac{1}{|\rt| \lqcd}+e\right)}\right],
\end{equation}
and we take $\qso^2=0.2$ GeV$^2$ and $\lqcd=0.241$ GeV. In Fig.~\ref{fig:fcBK_NLO} we show our results for the multiplicity as well as the NLO/LO ratio when using the unsubtracted~(\ref{eq:nc_unsub}) and the subtracted~(\ref{eq:nc_sub}) formulations for the $\nc$ terms. We see that these formulations are indeed equivalent and lead to positive results at all transverse momenta. The CXY approximation, on the contrary, leads to negative cross sections for $\ktt \gtrsim 10$ GeV.

The equivalence between the unsubtracted and subtracted formulations holds only if one uses the same value for $\as$ in the cross section and when solving the BK equation. Most LO studies taking into account running coupling corrections, such as Ref.~\cite{Lappi:2013zma}, use the Balitsky prescription~\cite{Balitsky:2006wa} for this. Being a coordinate-space prescription, this cannot be used exactly in our momentum space formulation. Trying to use BK solutions obtained with the Balitsky prescription together with a momentum-space formulation of the cross section leads to a breaking of the equivalence between~(\ref{eq:nc_unsub}) and~(\ref{eq:nc_sub}), with the subtracted version leading to negative cross sections at large transverse momenta~\cite{Ducloue:2017mpb}. A possible solution to this problem would be to perform the whole calculation in coordinate space, which allows to choose a running coupling prescription which matches the Balitsky one in the appropriate $\xi \to 1$ limit. However, as shown in Ref.~\cite{Ducloue:2017mpb}, a rather straightforward implementation of this approach leads to unphysical results, with a NLO result orders of magnitude larger than the LO one.

\section{Conclusions}

In this work we have shown that the formulation of the NLO cross section for single inclusive hadron production proposed in Ref.~\cite{Iancu:2016vyg} indeed leads to physical results at large transverse momenta for fixed values of the running coupling. Still, further developments will be needed before phenomenological studies will be possible. First, one should take into account the other ($q \to g$, $g \to q$ and $g \to g$) channels as well as the fragmentation functions, which should in principle be rather straightforward. Another issue is related to finding a running coupling scheme which would be consistent with previous studies of DIS. Finally, a complete NLO calculation should use dipole correlators obtained by solving the NLO BK equation~\cite{Balitsky:2008zza,Lappi:2015fma}, or at least a collinearly-resummed version of the LO equation~\cite{Iancu:2015vea,Iancu:2015joa} that can be made to include a large part of the NLO effects~\cite{Lappi:2016fmu}.

\section*{Acknowledgments} 
We thank E. Iancu  and D. Zaslavsky for discussions and H. Mäntysaari for providing his BK solutions. This work has been supported by the Academy of Finland, projects 267321, 273464 and 303756 and  by the European Research Council, grant
ERC-2015-CoG-681707.

\providecommand{\href}[2]{#2}\begingroup\raggedright\endgroup

\end{document}